\documentclass[%
 reprint,
 superscriptaddress,
 amsmath,amssymb,
 aps,
 pre,
]{revtex4-2}

\usepackage{graphicx}
\usepackage{dcolumn}
\usepackage{bm}
\usepackage{hyperref}
\usepackage[mathlines]{lineno}
\usepackage{color}
\usepackage{gensymb}
\usepackage[USenglish]{babel}

\newcommand{\ec}[1]{\textcolor{black}{#1}}

\begin{document}

\title{Rigid Base Biasing in Molecular Dynamics enables enhanced sampling of DNA conformations}

%
%

\author{Aderik Voorspoels}
\affiliation{Soft Matter and Biophysics, Department of Physics and Astronomy, KU Leuven}
\author{Jocelyne Vreede}
\affiliation{Van ’t Hoff Institute for Molecular Sciences, University of Amsterdam, 
Science Park 904, 1098 XH Amsterdam, the Netherlands}
\author{Enrico Carlon}
\affiliation{Soft Matter and Biophysics, Department of Physics and Astronomy, KU Leuven}

\begin{abstract}
All-atom simulations have become increasingly
popular to study conformational and dynamical
properties of nucleic acids as they are accurate and
provide high spatial and time resolutions. This high
resolution however comes at a heavy computational
cost 
\ec{and within the} time scales \ec{of simulations}
nucleic acids weakly fluctuate around their
ideal structure 
\ec{exploring}
a limited set of conformations. We introduce the
RBB-NA algorithm which is capable of controlling
rigid base parameters in all-atom simulations of
Nucleic Acids. With suitable biasing potentials
this algorithm can “force” a DNA or RNA molecule
to assume specific values of the six rotational (tilt,
roll, twist, buckle, propeller, opening) and/or the
six translational parameters (shift, slide, rise, shear,
stretch, stagger). The algorithm enables the
use of advanced sampling techniques to probe the
structure and dynamics of locally strongly deformed
Nucleic Acids. 
We
illustrate its performance showing
some examples in which DNA is strongly
twisted, bent or locally buckled.
In these examples RBB-NA reproduces well the
unconstrained simulations data 
\ec{and other} known features of DNA mechanics\ec{, but
it also allows one to explore the anharmonic behavior 
characterizing the mechanics of nucleic acids in
the high deformation regime.
}
\end{abstract}

\maketitle

\section{Introduction}
\label{sec:introduction}
All-atom molecular dynamics (MD) simulations have become standard tools to investigate 
the mechanical properties of DNA \cite{lank03,noy12,pasi14,vela20,walt20}. These properties 
are of high relevance in biological processes, as DNA is often physically deformed by 
proteins or other biomolecules (see Refs.~\cite{agga20,dohn21,mari21} for recent reviews 
on DNA mechanics).

DNA force fields have been constantly improved over the years using inputs from experiments 
and quantum mechanical calculations \cite{ivan16} and simulations have reached a high degree 
of sophistication.  At present DNA simulations have also some drawbacks. They are CPU-intensive 
and can thus sample molecular trajectories for 
only short time intervals ($\sim 1\, \mu s$) and short DNA sequences ($\sim 20-30$ base pairs). 
In these trajectories a DNA molecule fluctuates weakly around its ideal double helical
structure. To investigate strongly deformed conformations one has to resort to advanced 
sampling techniques. Many enhanced sampling techniques add potentials that constrain 
or drive the system along a predefined collective variable.
This collective variable typically involves several atoms and represents a relevant 
degree of freedom. In the past, umbrella sampling was used to induce strong 
bending by constraining the two ends of a linear DNA molecule \cite{Curu09}. 
While this approach allows one to compute the global bending free energy and to analyse
kinked DNA, it provides no control at the local scale, as the bending deformation gets 
distributed over several base pairs. 
Likewise, by applying a force or a torque at the two ends one can stretch or twist a DNA 
molecule, but this stretching or twisting is typically distributed over all base pairs 
\cite{mari17,shep22}. 
On a more local scale a different advanced sampling method, metadynamics along a 
path in a space spanned by several coordinates representing aspects of the DNA 
dynamics, was used to study Hoogsteen base pairing \cite{orti22}.

In this paper we introduce RBB-NA (rigid base biasing of nucleic acids), an algorithm 
which, by using suitable local constraints, is capable of deforming a DNA molecule by 
bending, stretching or twisting at the single base pair level. It enables using 
rigid base coordinates as the collective variable in advanced sampling techniques. 
Although we focus on a few illustrative examples on DNA, the algorithm also works 
with RNA.
Different constraints can be applied simultaneously (as e.g.\ bend and twist) either
to a single site or to multiple sites, contiguous or not, of a DNA sequence.
As such, RBB-NA is a very flexible tool to investigate structure, energetics and 
dynamics of highly deformed DNA molecules in MD simulations.
RBB-NA is available as a package 
in the Open Source Library PLUMED \cite{Trib14,plumed}, implemented as a collective variable.
PLUMED already contains much functionality to allow advanced sampling techniques, such as 
metadynamics, umbrella sampling and various path sampling techniques.


\section{MATERIALS AND METHODS}
\label{sec:Method}

\subsection{The RBB-NA algorithm}

\begin{figure}[t]
    \centering
    \includegraphics[width=\linewidth]{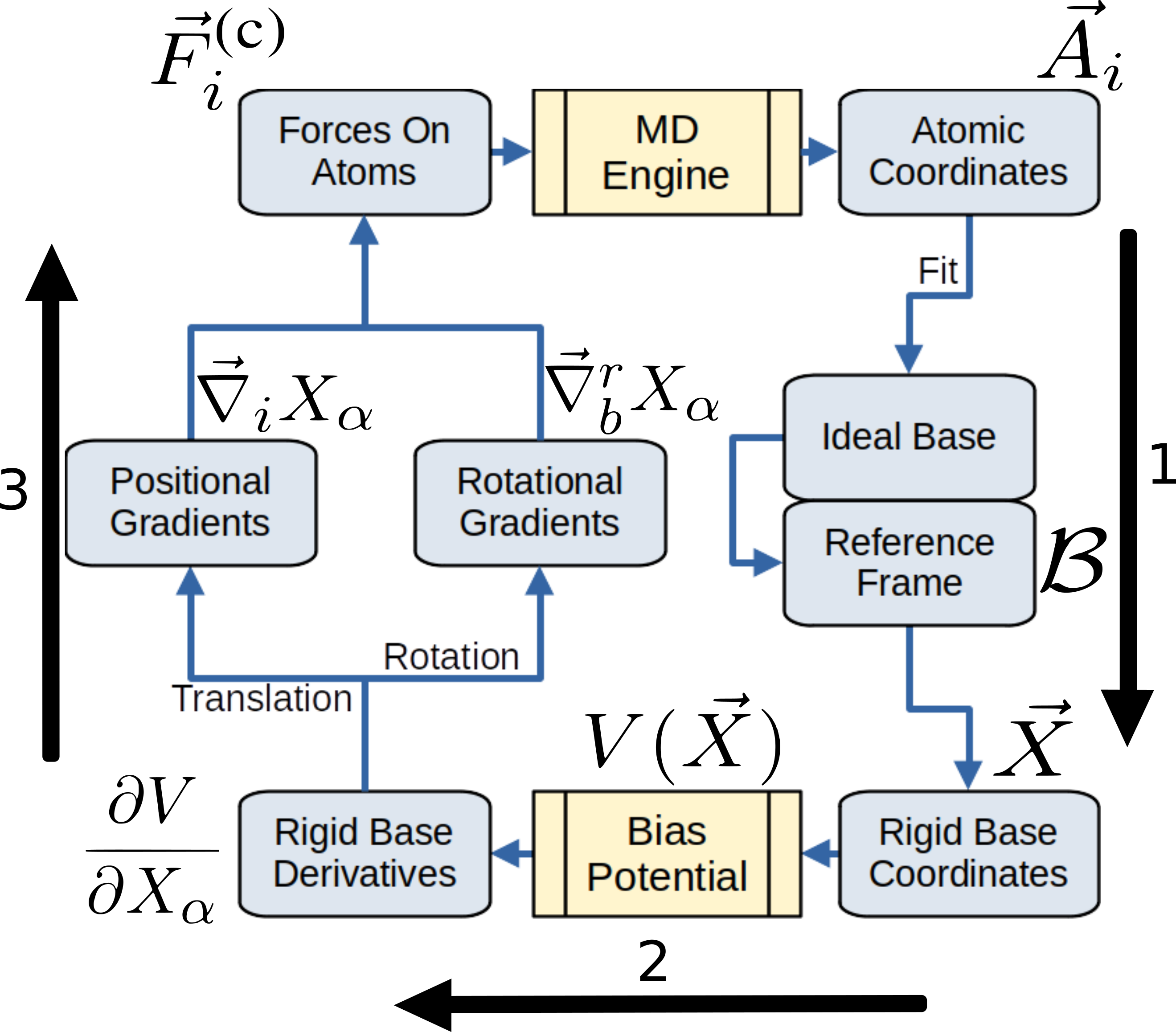}
    \caption{A flowchart showing the operations done by RBB-NA. The numbered arrows correspond 
    to (1) Mapping onto rigid base coordinates, as done in Curves+, (2) Introducing constraints 
    at rigid base level using the potential $V(\vec{X})$ (3) Calculating constraints forces on 
    atoms $\vec{F}_i^\text{(c)}$. This third step consists of translational and rotational 
    constraints. Total forces used in the MD integration step are the sum of constraints 
    ($\vec{F}_i^\text{(c)}$) and force fields contributions ($\vec{F}_i^\text{(ff)}$).
    Constraints are typically applied to a limited set of atoms in the bases, hence the
    running of the algorithm does not significantly slow down the running of the MD code.}
    \label{fig:flowChart}
\end{figure}

The aim of the algorithm is to compute suitable constraint forces $\vec{F}_i^\text{(c)}$ 
which act on a group of atoms ($i$ labels a given atom) and induce different bending, 
twist or stretching. The flowchart shown in Fig.~\ref{fig:flowChart} depicts the operations 
of RBB-NA. It essentially consists of three different steps (labeled as 1, 2 and 3 in 
Fig.~\ref{fig:flowChart}) which are schematically described in the rest of this Section.

\subsubsection{1. From atomic to rigid base coordinates.}

\begin{figure}[t]
        \centering
        \includegraphics[width=\linewidth]{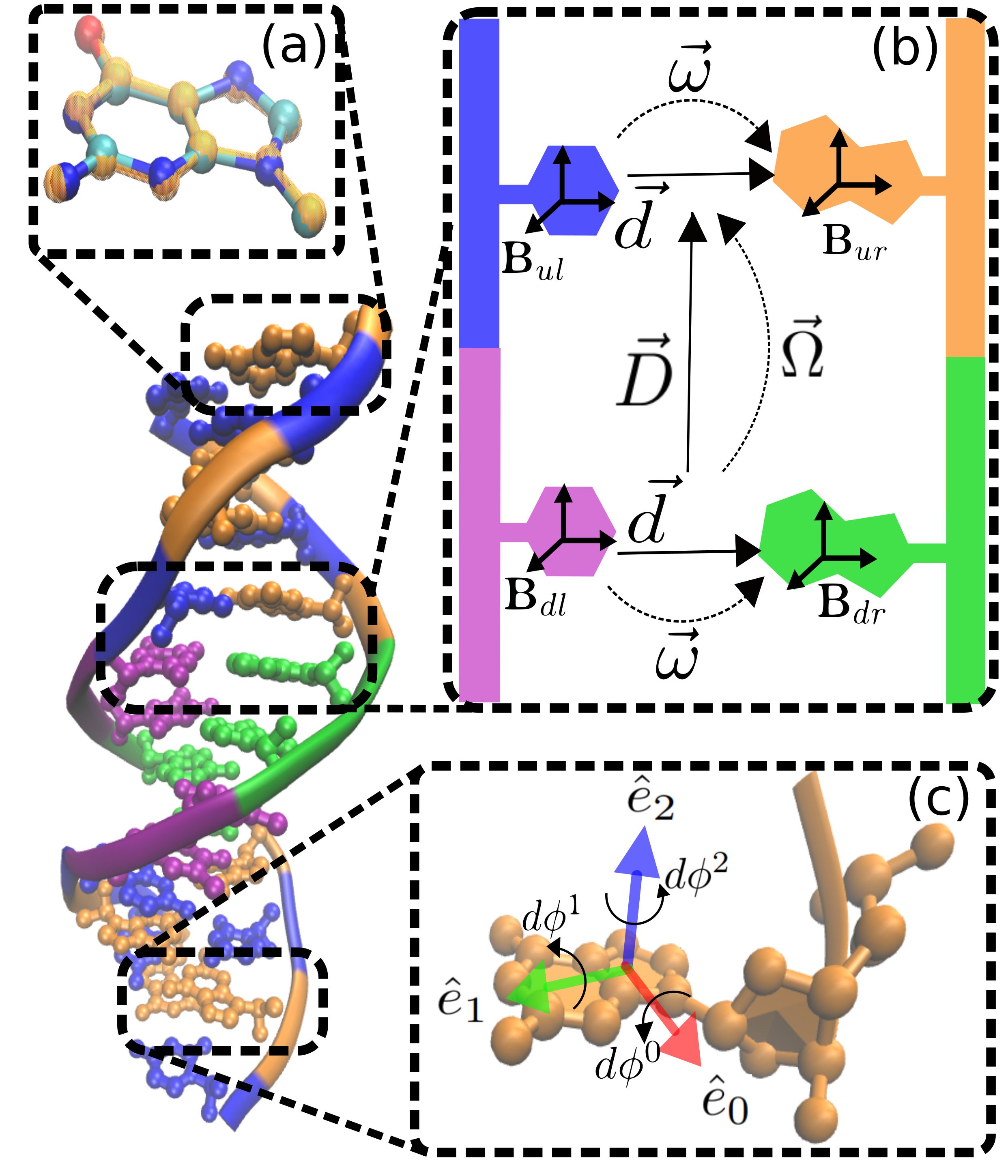}
        \caption{Illustration of different steps of the RBB-NA algorithm. (a) Mapping of atomic 
        coordinates of a base (colored) into an ideal base configuration (orange). (b) From the 
        reference frames ${\cal B}=(\mathbf{B},\vec{r})$ consisting each of an orthonormal triad 
        $\mathbf{B}$ and its origin $\vec{r}$ one calculates translational $(\vec d, \vec D)$ and 
        rotational $(\vec \omega, \vec \Omega)$ rigid base coordinates. $(\vec d, \vec \omega)$ are 
        intra-base coordinates, while $(\vec D, \vec \Omega)$ are inter-base coordinates. 
        (c) Rigid body rotations are used to calculate the constraint forces $\vec{F}^{(c)}$.
        Steps (a) and (b) are as in the Curves+ algorithm \cite{Lave09}.}
        \label{fig:stepDiagram}
    \end{figure}

This first step performs a mapping from atomic coordinates $\vec{A}_i$ to rigid base coordinates
for the bases to which the constraint is applied. Rigid base coordinates have long been a standard 
description of the double helix down to the base level \cite{olso01}. Software packages to extract 
rigid base coordinates, such as Curves+ \cite{Lave09} and x3dna \cite{x3dna}, were developed in the past. 
Usually these packages are used in a post-processing analysis step of MD trajectories.
In our algorithm we have implemented Curves+ to map atomic coordinates to rigid base coordinates 
during the simulation run. 
We briefly recall how Curves+ works. To obtain the rigid base coordinates one first associates to each 
base a reference frame ${\cal{B}} \equiv \{ \mathbf{B}, \vec r \}$ consisting of an orthonormal triad of 
unit vectors $\mathbf{B} = \left[ \hat{e}_{1}, \hat{e}_{2}, \hat{e}_{3}\right]$ (Fig.~\ref{fig:stepDiagram}) 
and their origin $\vec r$. 
This step conventionally involves fitting a set of ``ideal'' base coordinates to the actual atomic 
coordinates measured in simulation. To parametrize the translations and rotations connecting two 
complementary base frames on the opposite strands of a DNA molecule one uses the rigid base coordinates 
$\vec d$ (a vector whose components are referred to as shear, stretch and stagger in the DNA literature 
\cite{olso01}) and $\vec \omega$ (buckle, propeller and opening). The coordinates $\vec D$ (shift, slide 
and rise) and $\vec \Omega$ (tilt, roll and twist) parametrize translations and rotations between two 
consecutive base-pairs, see Fig.~\ref{fig:stepDiagram}. More details are given in SI.

\subsubsection{2. The constraint potential.}

For a sequence of $N$ base pairs one has $N$ base pair coordinates ($\vec{d}$, $\vec{\omega}$) 
and $N-1$ base pair steps coordinates ($\vec{D}$, $\vec{\Omega}$). We collect all rigid base 
coordinates into a single $12N-6$ dimensional vector
\begin{equation}
    \vec{X} \equiv (\vec \omega_1, \vec d_1,  \ldots \vec \omega_N, 
    \vec d_N, \vec\Omega_1, \vec D_1 \ldots \vec\Omega_{N-1}, \vec D_{N-1})
\end{equation}
In principle, the constraint potential is a generic function $V(\vec{X})$ of all these coordinates.
In practice, we fix constraints to a limited number of base pairs and rigid base coordinates, 
as shown in the examples discussed in Results.

\subsubsection{3. Translational and Rotational constraints.}

In order to implement the constraints in the MD simulation one needs to compute the 
forces acting on each particle. The constraint force on the $i$-th particle is given by
\begin{equation}
    \vec{F}_i^\text{(c)} = - \vec\nabla_i V(\vec X)
    = -\sum_\alpha \, \frac{\partial V}{\partial X_\alpha} \, \vec\nabla_i X_\alpha 
\label{Fc_simple}
\end{equation}
where the sum runs through all $12N-6$ components of the vector $\vec{X}$, $\alpha$ 
labels rigid base coordinates
and $\vec\nabla_i$ 
is the gradient with respect to the cartesian coordinates of the $i$-th particle.
The derivatives $\dfrac{\partial V}{\partial X_\alpha}$ are trivial to compute as the
function $V(\vec{X})$ is analytically simple (see examples in Results).
To compute $\vec\nabla_i X_\alpha$ one would need to know the function that maps
atomic coordinates into rigid base coordinates $\vec X$. However, the relation 
between the two sets of variables is complex, in part because it makes use of 
fitting procedures. Unfortunately these fitting methods make the mapping between 
non-differentiable. Moreover, the mapping is not one-to-one, as there are multiple 
atomic configurations that map to the same rigid base coordinate set.  
The approximation made here to resolve these issues, is to treat the bases 
as rigid when calculating and applying constraint forces. To incorporate the entire 
base into the computation of $\vec{X}$, and to reduce noise ideal base atomic coordinates 
are fitted to the measured ones. Afterwards, when forces are calculated, the measured 
coordinates are taken to represent a rigid body.

There are two different types of rigid base coordinates: those associated to translations 
($\vec d, \vec D$) and those associated to rotations ($\vec \omega, \vec \Omega$).
The calculation of $\vec\nabla_i X_\alpha$ for translations is trivial. This is just a
unit vector parallel to the component of the translation $X_\alpha$. 
For rotations the construction of $\vec\nabla_i X_\alpha$ is less straightforward.
It requires the computation of rotational gradients $\vec\nabla_{b}^{r} X_\alpha$ 
with respect to the orientation of the bases as a whole, from which torques can be calculated. 
Using rigid body dynamics these torques can then be converted to forces. Details of 
this procedure are given in SI.

Once the constraint forces $\vec{F}_i^\text{(c)}$ on all the atoms in the constrained 
bases are computed, we add to these the force fields contributions $\vec{F}_i^\text{(ff)}$ 
so the total force on the i-th atom is 
\begin{equation}
    \vec{F}_i^\text{tot} = \vec{F}_i^\text{(c)} + \vec{F}_i^\text{(ff)}
\end{equation}
This is used in the MD step to update positions and velocities from the current time $t$
to $t+\Delta t$. We note that in our scheme only atoms in bases are constrained, as these 
are the atoms used by Curves+ to calculates rigid base coordinates. Atoms in the backbone 
are not constrained.

\subsection{Simulations} 
\label{sec:simulationDetials}

\subsubsection{System Preparation.}
All simulations were done using version 2020.4 of Gromacs \cite{gromacs},
version 2.8.0 of PLUMED \cite{Trib14} and the Amberff99 parmbsc1 force field 
\cite{ivan16}. The RBB-NA algorithm has been implemented to work with Gromacs, 
however, because PLUMED is a highly portable plugin, it can be made to work 
with a multitude of MD engines after some minor adjustments.
Water was modelled using the TIP-3P model \cite{jorg83}, non-bonded 
interactions were cutoff at $1.0$ nm and PME Mesh Ewald interactions 
was used for electrostatics. All simulations started from the 
structure of the Drew-Dickerson (DD) dodecamer \cite{Dick89}, a 
self-complementary oligomer with sequence CGCGAATTCGCG which has been extensively 
investigated in prior experimental and computational studies (see 
e.g.\ Ref.~\cite{drsa13}). The DD sequence was placed into a dodecahedral 
box, leaving 2.0 nm on either side of the molecule, with periodic 
boundary conditions and solvated in a $150$ mM NaCl solution 
after which the overall charge in the system was neutralised.
This structure was energy minimised with a tolerance of $1000$~kJ/mol 
to make sure no overlap remained between solvent molecules and DNA. 
Subsequently, the molecule was equilibrated in the NVT ensemble for 
$100$~ps where temperature was kept at $300$~K using a velocity 
rescaling thermostat \cite{buss07} and then equilibrated for another $100$~ps 
in the NPT ensemble at the same temperature but using a Parrinello-Rahman 
barostat \cite{parr81} to fix the pressure at $1.0$~bar. 
These first equilibration simulations were performed using a $2$~fs time 
step in a leapfrog integrator, using LINCS \cite{Hess97} to constrain the 
covalent bonds involving hydrogen atoms.

\begin{figure*}[!t]
    \includegraphics[width=0.34\textwidth]{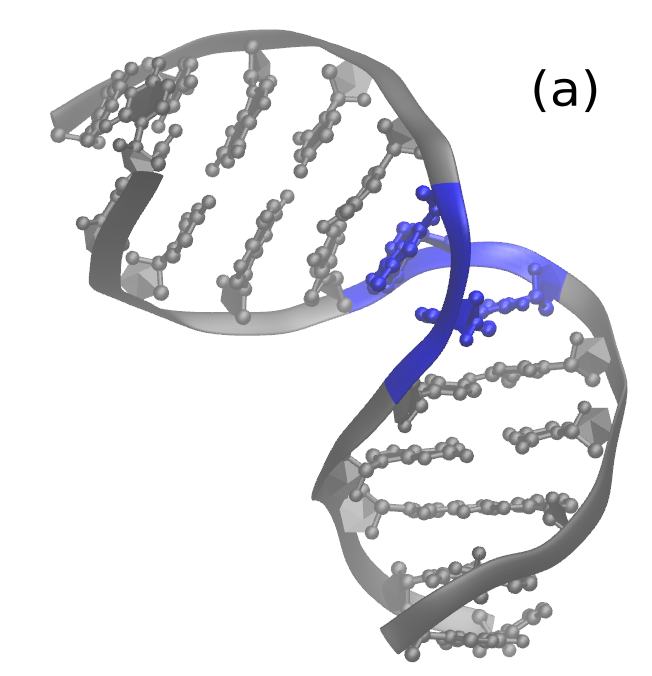}
    \hfill
    \includegraphics[width=0.62\textwidth]{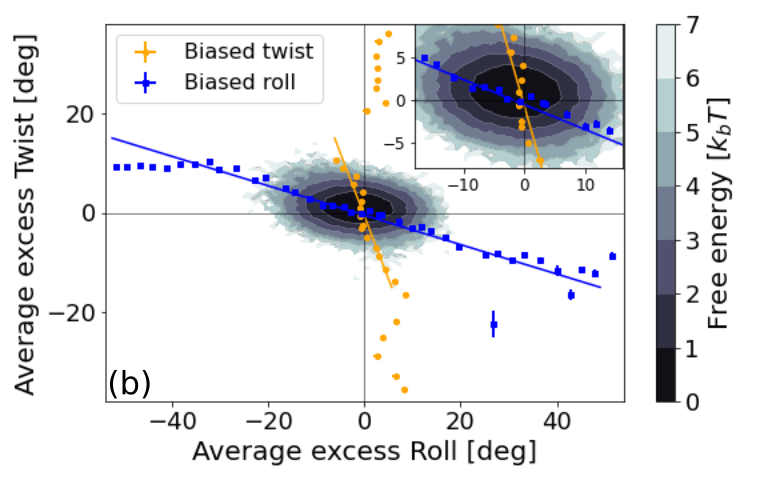}
    \caption{Simulations data obtained from the RBB-NA algorithm. (a) Snapshot of a 
    simulation in which a local bending is imposed by a constraint on the 
    roll $\Omega_2$ (via Eq.~\eqref{eq:Vroll} with with $K_r=1000$~kJ/mol and
    $\bar{\Omega}_2 = 1.0$~radian) in the central AT pair of the DD dodecamer. 
    The twist-roll coupling induces a bending of the molecule. (b) Colored circles: 
    average excess roll vs. average excess twist in simulations biasing the twist 
    via \eqref{eq:Vtwist} (blue circles) and the roll via Eq.~\eqref{eq:Vroll} 
    (orange circles). For small excess roll and twist the relation between these two
    variables is linear, as expected in the harmonic regime (solid lines, see SI for 
    details). In the simulations we used $K_t=K_r=1000$~kJ/mol.
    The contour plot, shown in details in the inset, is a free energy calculation from 
    unbiased MD simulations.}
     \label{fig:DeformAndCouple}
    \end{figure*}

\subsubsection{Production Runs.}
A reference unconstrained MD simulation of $100$~ns was performed with the 
same settings as the NPT equilibration. Snapshots were stored every $1$~ps 
and collective variables were printed with the same frequency.
Subsequently three sets of simulations were done using the RBB-NA algorithm. 
All three were umbrella sampling runs which will be analysed for different 
applications in the results section\ref{sec:Examples}. 
In the runs we used three different type of constraints induced by
parabolic potentials
\begin{equation}
    V_\alpha = \frac{K_\alpha}{2} \left( X_\alpha - \bar{X}_\alpha \right)^2
\label{eq:Vgeneric}
\end{equation}
where $X_\alpha$ is either Roll, Twist or Buckle. The Roll and Twist constraints
were applied on the central base pair step (AT) of the DD sequence. The Buckle
constraint was applied on the second GC pair of the DD sequence. In all constrained
simulations we set the spring constant to $K_\alpha=1000$~kJ/mol.


For the case of the Roll simulation run the system was first pulled 
from its equilibrium configuration. This was done by running molecular dynamics 
in consecutive $100$~ps pulling windows with the same settings as the 
NPT equilibration but decreasing the time step to $0.2$~fs.
In these runs the average Roll, $\bar{X}_\alpha$ in the potential \eqref{eq:Vgeneric}, 
was increased(or decreased) by $0.01$~radians (starting from a zero Roll average) 
every pulling window. 
Here a maximum of $1.0$~radians and a minimum of $-1.0$~radians were reached. 
The system resulting from every fifth pulling window (ie with $\bar{X}_\alpha$ 
increasing in steps of $0.05$ radians) was then used to start a $10$~ns production run.
Here again the same settings as the NPT equilibration were used now setting the 
time step back to $2.0$~fs. This resulted in a total of $41$ windows of $10$~ns with 
the imposed average Roll spanning from $1.0$ to $-1.0$~radians.

After the production runs for umbrella sampling on Roll a similar procedure was 
carried out for Twist in the central step. 
Again the system was pulled from the equilibrium configuration in $100$~ps windows 
(with the same settings as the NPT equilibration but decreasing the time step to $0.2$~fs), 
using a potential of the shape \eqref{eq:Vgeneric}, while increasing  and decreasing 
the imposed average $\bar{X}_\alpha$ in steps of $0.01$~radians.
Notably for these Twist constraining simulation the intrinsic twist of DNA was taken 
into account by not starting from a zero twist average.
Instead $0.61$~radians was the staring average.
The maximum imposed twist reached was $1.31$~radians, the minimum was $-0.09$~radians.
Like for Roll every fifth window was used to start a production run of $10$~ns, with 
the same settings as the NPT equilibration restoring the time step to $2$~fs.
This resulted in $29$ windows spanning an imposed average twist from $-0.09$ to $1.31$~radians.

Finally for Buckle of the second GA pair, again the same procedure was used.
The system was pulled in $100$~ps  windows (with the same settings as the NPT 
equilibration but decreasing the time step to $0.2$~fs) with the buckle changing 
form $0.0$~radians in $0.02$~radian steps.
The maximum reached here was $1.3$~radians, the minimum was $-1.3$~radians.
The system resulting from every fifth pulling window (ie with the imposed average 
Buckle increasing in steps of $0.1$~radians) was then used to start a $10$~ns 
production run. This resulted in $17$ windows.

The results in these simulations were analysed to be used in three different examples: 
showing coupling between Twist and roll, characterising the free energy landscape and 
illustrating interactions between neighboring sites in DNA.

\subsection{Weighted Histogram Analysis} 
\label{sec:WHAM}
A detailed mathematical description of 
umbrella sampling and the Weighted Histogram Analysis Method (WHAM) that is used to 
reconstruct the free energy landscape has been given by J. Kaestner \cite{Kast11}. 
In short the method entails constraining the conformation space of a DNA molecule by 
imposing a steep potential, which is usually quadratic in the desired coordinate, 
as in the examples \eqref{eq:Vtwist} and \eqref{eq:Vroll}. By changing the location 
of the minimum of the constraint potential (e.g.\ by varying $\bar{\Omega}_3$ and 
$\bar{\Omega}_2$ in \eqref{eq:Vtwist} and \eqref{eq:Vroll}) one can then control 
which region of the conformation space is sampled. If the sampled regions partially 
overlap, the relative free energy landscape over all regions can be reconstructed 
afterwards using WHAM analysis. This involves an iterative procedure\cite{Kast11} 
which here was considered to have converged if the integral of the difference between 
the free energies of the previous an current iterations was below $10^{-7}k_BT$.

\section{RESULTS}
\label{sec:Examples}
 
The RBB-NA algorithm was developed to enable umbrella sampling to characterise 
the free energy landscape by constraining local rigid base coordinates of DNA.
However, the same algorithm can also be used in any other type of enhanced 
sampling applications, such as metadynamics or steered MD.
We conducted three sets of umbrella sampling simulations, one imposing 
a local twist, one imposing a local roll and one imposing a local Buckle. 
From these simulations we present three different types of analysis which 
illustrate the functioning of the algorithm.
In the first analysis, we explore how imposing a local roll or twist affects 
other rigid base coordinates. We find that a bias in the twist induces a 
non-vanishing excess roll and, vice versa, a bias in the roll induces a 
non-vanishing excess twist. This is a clear demonstration of the effect of 
twist-roll coupling \cite{mark94}. In the second analysis, we compute free 
energy profiles along the imposed twist or roll coordinate. Here our 
analysis points to anharmonic behavior. Finally, the third analysis 
illustrates the effect of a local bias on base pairs flanking the 
constrained site.

\subsection{Twist-Roll Coupling}\label{sec:Coupling}

Figure \ref{fig:DeformAndCouple}(a) shows a snapshot of an MD simulation in 
which an average excess roll of $\sim60^\circ$ is imposed on the central 
base pair step (AT) of the DD sequence for which a local constraint potential 
of parabolic type was used
\begin{equation}
    V_\text{roll} = \frac{K_r}{2} \left( \Omega_2 - \bar{\Omega}_2 \right)^2
\label{eq:Vroll}
\end{equation}
(we use $\Omega_1$, $\Omega_2$ and $\Omega_3$ to indicate the excess tilt, roll 
and twist). As shown in the snapshot of Fig.~\ref{fig:DeformAndCouple}(a) the 
bending induces a remarkably strong twisting deformation as well, due to the 
effect of the twist-roll coupling \cite{mark94}.
In homogeneous DNA models such interaction is described by a term of the 
type $G \Omega_2 \Omega_3$, with $G$ the twist-roll coupling. Twist-roll 
coupling and its influence on the mechanical properties of DNA has been 
discussed in the recent literature \cite{skor18,cara19,nomi19a}.

A more structured and in depth view of this coupling is given in 
Fig.~\ref{fig:DeformAndCouple}(b),  which shows the average excess 
roll vs. the average excess twist for different simulations biasing 
in the roll via Eq.~\eqref{eq:Vroll}, as well as for different simulations
biasing in the twist, using a constraint potential of the type 
\begin{equation}
    V_\text{twist} = \frac{K_t}{2} \left( \Omega_3 - \bar{\Omega}_3 \right)^2
\label{eq:Vtwist}
\end{equation}
with varying values of $\bar{\Omega}_3$. This constraint imposes indeed a finite excess twist, 
but also a roll (Fig.~\ref{fig:DeformAndCouple}(b)). 
Note that, for weak deformations, excess twist and roll induced by the biasing 
potentials \eqref{eq:Vroll} and \eqref{eq:Vtwist} are linearly correlated. 
Moreover, they follow lines with different slopes (orange and blue lines in 
Fig.~\ref{fig:DeformAndCouple}(b)), a behavior which can be explained by a simple 
harmonic model calculation, see SI.
Induced excess Twist and Roll also have opposite signs, as expected for a positive 
twist-roll coupling $G>0$. The contour lines in Fig.~\ref{fig:DeformAndCouple}(b) 
show the free energy calculated by Boltzmann inversion ($F=-k_B T \ln(P)$,
with $P$ the equilibrium probability distribution) from the sampling of unconstrained 
MD simulations. The RBB-NA algorithm generates DNA conformations with roll and twist 
angles which greatly exceed those generated by unbiased simulations.
\ec{We note that in the high deformation regime the excess twist and roll are more 
weakly coupled to each other than in the harmonic regime. For instance, biasing the excess twist 
above $20^\circ$ does not produce a significant roll and a roll bias below
$-30^\circ$ does not lead to an increase of the twist, see 
Fig.~\ref{fig:DeformAndCouple}(b). While the mechanics of DNA in the weak deformations
regime has been intensively studied \cite{pasi14,vela20}, the strong deformation regime
is still largely unexplored. This regime is now within the reach of the RBB-NA algorithm.}

\subsection{Free energies from umbrella sampling} 
\label{sec:Umbrella}

\begin{figure}[t]
    \includegraphics[width=0.45\textwidth]{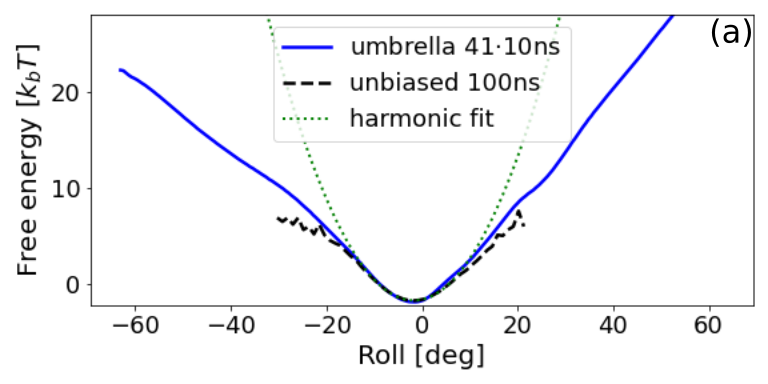}
    \includegraphics[width=0.45\textwidth]{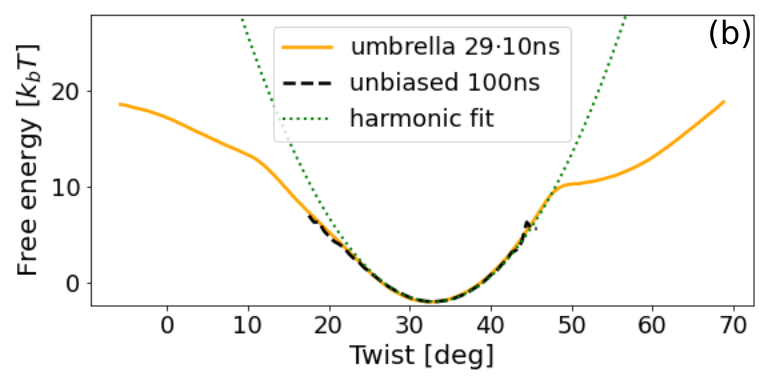}
       \caption{
      Free energy of roll (a) and twist (b) deformation of the central (AT) step
      of the DD dodecamer. Solid orange line: Umbrella sampling with the CRT-NA algorithm. 
      Dashed lines: Sampling from unconstrained simulations. The latter can generate only
      small deformations around the minimal free energy state. The umbrella sampling reproduces
      very well the unconstrained simulations data. Dotted lines: parabolic fits to the 
      unconstrained simulation. Both examples show deviations from the ``ideal'' parabolic 
      behavior indicating that highly bent or highly twisted deformations are 
      energetically less costly than a quadratic model would predict. Both data show
      asymmetries with respect to positive or negative excess roll and twist.
      }
    \label{fig:Umbrella}
    \end{figure}

The tight control over rotation angles of the RBB-NA algorithm can be used for 
umbrella sampling to obtain free energies. 
Figure~\ref{fig:Umbrella} shows a comparison between free energies obtained by 
unconstrained simulations (dashed lines) and with the RBB-NA algorithm via 
umbrella sampling (solid lines). The latter are calculated using the constraints
\eqref{eq:Vtwist} and \eqref{eq:Vroll}. A much wider range of twist and roll angles 
is available via umbrella sampling. For roll, Fig.~\ref{fig:Umbrella}(a), beyond 
a narrow interval around the free energy minimum, the landscape no longer follows 
a parabolic shape. 
The roll data instead suggest a behavior reminiscent of the linear sub-elastic chain
(LSEC) model, which was introduced to explain anomalous high bending of DNA as
observed in Atomic Force Microscopy (AFM) experiments \cite{wigg06} (we note  
that AFM data analysis has lead to some debate and to different conclusions about 
the existence of an actual anomalous bending behavior \cite{mazu14a}). The LSEC
model posits that the bending free energy of DNA is quadratic for low bending 
angles, while it is linear beyond a given threshold \cite{wigg06}. This implies 
that large bend angles are less costly than the quadratic model would predict. 
A free energy consistent with the LSEC model was also observed in prior all-atom 
simulations \cite{Curu09}, where a short DNA sequence was forced to bend by applying 
a constraint at its two ends. Such constraint induces bending which is distributed 
over several nucleotides, while the RBB-NA constraint of Fig.~\ref{fig:Umbrella} 
acts only on two adjacent base pairs (twist is an inter-base coordinate).
The control of rigid base coordinates offered by the RBB-NA algorithm will hopefully
provide more insights on the microscopic origin of the LSEC behavior.

The free energy landscape for twist is shown in Fig.~\ref{fig:Umbrella}(b). Again
we note a good overlap between unconstrained and umbrella sampling simulations.
Compared to the roll data, the twist follows a harmonic behavior for a wider
range of angles. For high over and undertwisting we observe deviations from the 
parabolic free energy profile, which indicate that highly deformed twisted 
conformations are more likely than a quadratic model would predict. The deviations 
from the harmonic behavior are however different than in roll, showing more abrupt 
transitions.
The overtwisted regime closely follows the quadratic shape predicted by the 
TWLC until a total twist of $\approx 48^\circ$ is reached.
There, the landscape shows a sharp transition which, upon close inspection 
of the trajectories, can be linked to a structural change involving partial 
breaking of the hydrogen bonds. 
In the undertwisted regime one can also note an abrupt change of behavior
in the free energy around a total twist of $15^\circ$, although less sharp 
than in the overtwisted case. We note that here we have sampled free energies
for roll and twist as induced by biasing via Eqs.~\eqref{eq:Vtwist} and 
\eqref{eq:Vroll}. This analysis provides just a "one-dimensional" projection 
of the two dimensional free energy $f(\Omega_2,\Omega_3)$. The full calculation
of this free energy will be presented elsewhere.

\subsection{Effect of a local bias on neighboring sites} 
\label{sec:Decay}
 
Several studies have observed correlations between rigid base coordinates
separated by a few nucleotides along the sequence, indicating the existence 
of couplings between distal sites \cite{lank03,esla11,noy12}.
A recent analysis quantified these non-local couplings for tilt, roll 
and twist both for DNA \cite{skor21} and dsRNA \cite{sege22}. Very similar 
couplings were found in these two molecules \cite{sege22}. Here we analyze 
structural deformations along the whole sequence of a DD dodecamer 
induced by local perturbations at a specific site.

\begin{figure}
\centering
\includegraphics[width=0.48\textwidth]{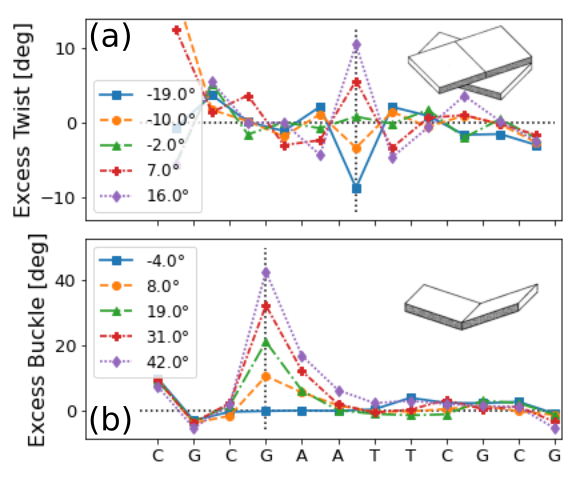}
\includegraphics[width=0.48\textwidth]{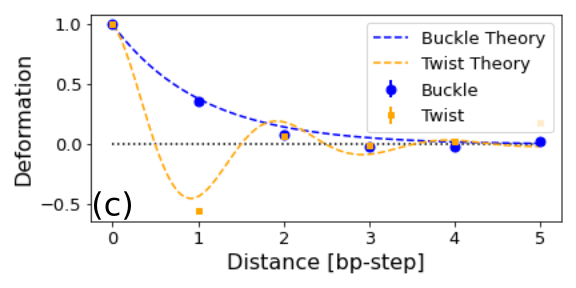}
\caption{
(a) Average excess twist induced along the molecule by a twist deformation 
imposed by a constraint of type \eqref{eq:Vtwist} in the central AT step of the DD 
dodecamer. The twist has an oscillatory decay as predicted by a twist 
free energy of type \eqref{FreeEnergy_Twist} with $\tilde{C}'>0$. Note
that Curves+ reports an anomalous high twist at the left end of the 
sequence, presumably due to some end effects. (b) Average excess buckle
induced by a constraint of type \eqref{eq:Vbuckle} applied to the
second G nucleotide from the left. (c) Average decays of twist and buckle from
MD simulations (symbols) and predictions from continuum non-local Twistable Wormlike
chain model \cite{skor21,sege22}.}
\label{fig:decay}
\end{figure}
    
Figure~\ref{fig:decay}(a) shows the average excess twist in the DD 
sequence upon imposing an excess twist via a potential
of the type \eqref{eq:Vtwist} on the central AT site, with the different
lines referring to different strengths of the perturbation. 
We note that the perturbation produces an oscillating decay of the
twist. This is in line with the predictions of non-local models 
of DNA and dsRNA elasticity \cite{skor21,sege22}.
This behavior can be understood by a minimal (homogenous) model of twist free 
energy $E_\text{twist}$ of the following type
\begin{equation}
    E_\text{twist} = \frac{1}{2} \sum_n \left[ \tilde{C} \Omega_3^2(n) 
    + \tilde{C}' \Omega_3(n) \Omega_3(n+1) 
    \right]
    \label{FreeEnergy_Twist}
\end{equation}
where the sum runs through all the sites and $\tilde{C}$ and $\tilde{C}'$
are the on-site and next-neighbor torsional stiffnesses and $\Omega_3(n)$
is the excess twist at site $n$. For simplicity we have limited the off-site 
interactions to a coupling between consecutive sites $n$ and $n \pm 1$, but 
this argument can be easily extended to interactions coupling more distant
sites, as $n$ and $n+2$ etc\ldots \cite{skor21}. 
Prior work showed that both DNA and dsRNA are characterized by $\tilde{C}'>0$ 
\cite{sege22}, while stability requires that $\tilde{C}>0$. Let us consider 
a perturbation of type \eqref{eq:Vtwist} enforcing $\langle \Omega_3(k) \rangle > 0$ 
at a given site $k$. As $\tilde{C}'>0$ energy minimization induces 
$\langle \Omega_3(k \pm 1) \rangle < 0$ at the two flanking sites. This 
ultimately produces an alternating decaying profile of over- and undertwist.

Finally, we used RBB-NA to induce a buckle ($\omega_1$) which is an intra 
base-pair parameter describing the mutual orientation of two complementary 
bases in the same pair. Buckle is the angle formed by a rotation around 
the pair short axis. A non-zero buckle was induced using the following constraint
\begin{equation}
    V_\text{buckle} = \frac{K}{2} \left( \omega_1  - \bar{\omega}_1 \right)^2
    \label{eq:Vbuckle}
\end{equation}
Instead of an alternating pattern buckle shows a monotonic decay, see 
Fig.~\ref{fig:decay}(b). This type of behavior can be generated by a 
buckle free energy of the same form as Eq.~\eqref{FreeEnergy_Twist}, but 
with a negative off-site stiffness (corresponding to $\tilde{C}'<0$).
We note that the decay of the buckle is asymmetric, which may be
due to sequence or end effects.

Figure \ref{fig:decay}(c) compares the results of simulations directly 
to the theory of non-local DNA elasticity developed in \cite{skor21}. 
For twist, good agreement is found between the observed oscillatory 
decay and theoretical predictions (dashed lines).
Buckle, which was not included in previous studies, also decays 
exponentially along the chain as a result of a local perturbation, 
a clear signature of an off-site (non local) coupling.
Parameters needed to determine the theoretical decay profile, 
local stiffness and off site couplings, were determined using 
the same reference simulation used in the umbrella sampling simulations.

\section{DISCUSSION}
\label{sec:conclusions}
    
We have developed an algorithm (RBB-NA) which allows one to control
rigid base parameters in DNA. In RBB-NA suitable bias potentials are
used to impose specific values to one (or more than one) of the twelve 
rigid base parameters tilt, roll, twist, buckle, propeller, opening, 
shift, slide, rise, shear, stretch and stagger. The algorithm uses rigid 
body dynamics to translate potentials on complex collective variables 
into constraining forces which act on atoms during simulations. RBB-NA
can thus generate highly deformed DNA in a controlled manner at the local 
scale, e.g.\ at the base pair level. We believe that the algorithm offers 
an interesting new tool to explore the dynamics of rare events and to map 
the full free energy landscape, by using advanced sampling techniques, 
well-beyond what unconstrained simulations can reach.

After giving a schematic overview of the working of RBB-NA (see SI for more
mathematical details), we have discussed three examples. The first one shows 
how a bias in the twist actually induces an excess roll as well, and vice-versa, 
a bias in the roll induces a non-zero excess twist. This is an illustration 
of the twist-roll coupling. Although this coupling is well-known in DNA 
\cite{mark94} and also studied at length in the recent literature 
\cite{skor18,cara19,nomi19a}, the fact that we can observe its effect in 
RBB-NA is an indication that the algorithm works correctly. We find indeed 
that a positive excess twist bias induces a negative roll and  vice-versa, 
as expected in the case of positive twist-roll coupling constant $G>0$, 
in agreement with prior studies \cite{olso01}.
Apparent from these results is also the opportunity RBB-NA provides to study 
the structure of highly deformed conformations of DNA, such as the snapshot 
shown in Fig.~\ref{fig:DeformAndCouple}(a). Correspondingly one can see from 
Fig.~\ref{fig:DeformAndCouple}(b), comparing the area accessible in unconstrained 
sampling to the reach of the simulations using RBB-NA, the breadth of conformations 
which can now be accessed in a highly controlled manner.

The second example shows free energy calculations where RBB-NA was used 
to perform umbrella sampling simulations. We considered again roll and twist 
deformations as illustrative. The results of these simulations were analyzed using WHAM.
Notable here is that this relatively straightforward application of the algorithm 
already reveals some interesting features in the presented free energy landscape.
Especially interesting is the LSEC behaviour of Roll.
This result is in line with previous experimental \cite{wigg06} and computational
\cite{Curu09} findings. Here, however, we can for the first time attribute, the 
previously observed behaviour, to the local basepair step scale. As such RBB-NA is 
a promising tool that could be used in future studies to uncover the precise origin 
of this anharmonic behaviour. Moreover RBB-NA can be used in umbrella sampling 
studies to identify other relevant and interesting features in the free energy 
landscape of dsDNA like was done for twist. While here only ``one-dimensional'' 
projections of the entire free energy landscape have been discussed. RBB-NA might 
well be used to explore and map the full 12 dimensional free energy landscape 
or other interesting subsets thereof. This complete free energy landscape is likely 
to hold features (barriers, dips, local minima, ...) which are important for the 
biological functioning of DNA.
 
Finally, in a third example we investigated how a local perturbation propagates 
to neighboring sites. Here results were presented for twist and buckling deformations 
which show two different types of spatial propagation: twist decays through damped 
oscillations, while buckle shows a monotonic decay. This behavior can be attributed to 
the different signs of the non-local coupling terms, as discussed in \cite{skor21}.

We believe that the possibility of deforming DNA (or RNA) at the local scale 
will provide 
several new insights on the structural and dynamical properties of 
Nucleic Acids. 
In addition, the possibilities of computing free energies in the high deformation
regime, beyond the harmonic regime will lead to an improved parametrization of
rigid base coarse-grained models of DNA \cite{walt20,oroz08}.

\subsection{Availability}

The RBB-NA code is made available through github: 
\href{https://github.com/AderikVoorspoels/RBB-NA.git}{\url{https://github.com/AderikVoorspoels/RBB-NA.git}}.

\section*{Acknowledgements} \label{sec:acknowledgements}

We acknowledge interesting discussions with Midas Segers and
Enrico Skoruppa.



%

\vfill\eject

\renewcommand{\thesection}{S\arabic{section}} 
\setcounter{section}{0}
\renewcommand{\theequation}{S\arabic{equation}} 
\setcounter{equation}{0}
\renewcommand{\thefigure}{S\arabic{figure}} 
\setcounter{figure}{0}

\section*{SUPPLEMENTAL INFORMATION}

    \begin{figure}[t]
        \centering
        \includegraphics[width=\linewidth]{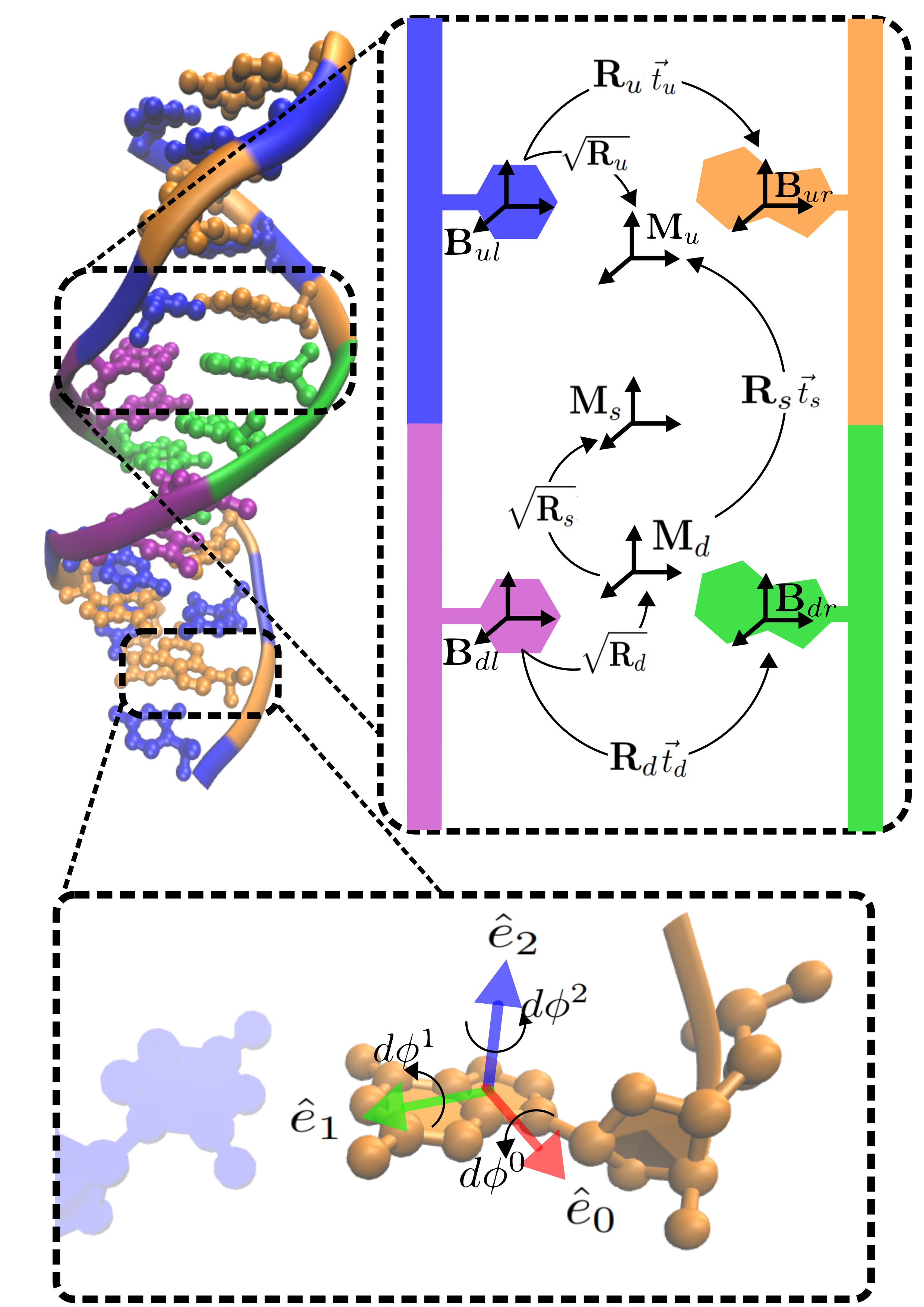}
        \caption{A strand of DNA. Attached to the right is a diagram showing 
        the relation between different rotations and reference triad. Attached 
        below is an illustration of the unit vectors forming a base triad and 
        the action small rotations have on them}
        \label{fig:DetailedDiagram}
    \end{figure}

\section{Mathematical derivations in the algorithm} \label{sec:MethodLong}

The purpose of the RBB-NA algorithm is to bias the rotational and 
translational rigid base coordinates $\vec{X}$ of a DNA (or RNA) strand, 
given the representation of this strand as a series of bases $\mathcal{B}_{b}$.
Here $b$ ranges from $0$ to $2N$ with $N$ being the length of the strand 
in question in base pairs. Notably any base will during the simulation be 
represented by a collection of atom positions $\left\{\vec{A}_i \right\}$.
It is from these atom positions that the calculation of the collective 
variables must start, and it is on these atoms that in the end forces can 
be applied using Eq.~\eqref{Fc_simple}.
For sake of completeness and to introduce the required notations and concepts 
this section begins with a description of the curves algorithm developed 
by Lavery et al. \cite{Lave09} which is used to calculate the collective 
variables. This algorithm has been re-implemented in RBB-NA to fit with 
the PLUMED code base.

\subsection{Curves in brief}

The standard choice in the Curves+ software is to first fit a set of 
ideal base coordinates $\left\{ \vec{A}_{i}^{*} \right\}$ to the 
measured ones by finding the appropriate rotation in a procedure 
described by McLachlan \cite{McLa79}. 
This is usually done to reduce the influence of fluctuations within 
a base on the collective variables, thus reducing the noise.
For the purposes of this work this step is useful as it incorporates 
all heavy atoms in a given base into the calculations of the collective 
variables. As such it is this step that facilitates both treating the 
entire base as a rigid body as well as preventing forces from 
interfering with fluctuations in the bases themselves.
    
After the optional fitting the curves algorithm starts by defining 
reference frames $\mathcal{B}_b$ for each base consisting of 
a triad of orthonormal vectors $\mathbf{B}_b$ and a reference 
point $\vec{r}_b$ where this frame is to be attached \cite{Lave09}. 
Here $\mathbf{B}_b$ can be seen as a rotation matrix
\begin{equation}
    \mathbf{B}_b = \left[ \hat{e}_{1}, \hat{e}_{2}, \hat{e}_{3} \right].
\end{equation}
where the vectors $\hat{e}_{1}, \hat{e}_{2}, \hat{e}_{3}$ collectively 
form the base triad with $\hat{e}_{1}$ pointing into the major groove, 
$\hat{e}_{2}$ connecting the backbones and $\hat{e}_{3}$ being normal 
to the base plane.
    
To keep track of the different frames and many variables needed in the 
calculations that follow we include here a diagram (\ref{fig:DetailedDiagram}).
On this diagram the reference frames involved in the calculation of rigid 
base coordinates in this particular step are labelled according to their 
position in relation to the step. We denote all frames in the upward direction 
from the step with a subscript `$u$', The frames in the downward direction 
are denoted similarly with subscript `$d$'. 
For frames on the left strand we add `$l$' and for those on the right  `$r$'.
    
When the frames are constructed one defines the transformation from one base 
to the opposing one, say in the diagram, \eqref{fig:DetailedDiagram} from 
$\mathcal{B}_{ul}$ to $\mathcal{B}_{ur}$, by rotation matrix $\mathbf{R}_u$ 
and translation $\vec{t}_u$ which are defined as
\begin{equation}
    \mathbf{R}_{u}=\mathbf{B}_{ur}\mathbf{B}_{ul}^{T}, 
    \quad \vec{t}_{u} = \vec{r}_{ur} - \vec{r}_{ul}.
\label{eqn:transformDefinitions}
\end{equation}
In addition to these transformations a frame in which to express 
them is needed. This is given by the midframe $\mathcal{M}_{u}$ of 
this particular base pair for which the triad can be written as
\begin{equation}
    \mathbf{M}_{u} = \sqrt{\mathbf{R}_{u}}\mathbf{B}_{ul} =  
    \sqrt{\mathbf{R}_{u}^{T}}\mathbf{B}_{ur}.
    \label{eqn:midpoint}
\end{equation}
Here the square-root $\sqrt{\mathbf{R}}$ is a short notation for 
$\exp{\frac{1}{2}\log{\mathbf{R}}}$ which yields a rotation about 
the same axis by half the angle.
    
In practice the midframe is not found by explicit calculation of 
the square root but simply by rotating the three vectors in 
$\mathbf{B}_{ul}$ around the axis of rotation of $\mathbf{R}_{u}$ 
by half its rotation angle. 
For any rotation matrix the angle of rotation $\Theta(\mathbf{R})$ 
can be found by 
\begin{equation}
    \Theta(\mathbf{R}) = \arccos{ \frac{1}{2}\left(\text{Tr}(\mathbf{R}) - 1 \right) }, 
    \label{eqn:MatrixToAngle}
\end{equation}
and subsequently the axis of rotation $\widehat{K}(\mathbf{R})$ is given as
\begin{equation}
    \widehat{K}(\mathbf{R}) = \frac{1}{2 \sin{\Theta(\mathbf{R})}} 
    \begin{bmatrix} 
    R_{23}-R_{32}\\ R_{31}-R_{13}\\R_{12}-R_{21}
    \end{bmatrix}. 
\label{eqn:MatrixToAxis}
\end{equation}
This conversion of rotation matrices to axis angle representation 
also allows us to define the rotational collective variables of a base pair as
\begin{equation}
    \vec{\omega}^{T}_{u} = \Theta(\mathbf{R}_{u}) (\widehat{K}(\mathbf{R}_{u})^{T} \cdot \mathbf{M}_{u}).
    \label{eqn:omegaDef}
\end{equation}
Similarly the translational collective variables of the basepair are defined by 
\begin{equation}
    \vec{d}^{T}_{u} = \vec{t}_{u}^{T} \cdot \mathbf{M}_{u}.
    \label{eqn:SDef}
\end{equation}
After the calculation of the basepair collective variables the 
algorithm largely repeats itself. The transformation from one 
pair to the next, say from $\mathcal{M}_d$ to $\mathcal{M}_u$ 
on the diagram, is now given by a rotation $\mathbf{R}_s$ and a translation 
$\vec{t}_{s}$ which are defined by
\begin{equation}
    \mathbf{R}_{s}=\mathbf{M}_{u}\mathbf{M}_{d}^{T}, \quad 
    \vec{t}_{s} = \frac{1}{2}(\vec{r}_{ur} + \vec{r}_{ul}) - 
    \frac{1}{2}(\vec{r}_{dr} + \vec{r}_{dl}).
\end{equation}
Like before a midframe is then defined by half rotation:
\begin{equation}
    \mathbf{M}_{s} = \sqrt{\mathbf{R}_{s}}\mathbf{M}_{d} =  
    \sqrt{\mathbf{R}_{s}^{T}}\mathbf{M}_{u},
\end{equation}
and finally the transformations are expressed in this midframe
\begin{eqnarray}
    \vec{\Omega}^{T}_{u} &=& \Theta(\mathbf{R}_{s}) 
    (\widehat{K}(\mathbf{R}_{s})^{T} \cdot \mathbf{M}_{s})
    \label{eqn:OmegaDef}  \\
    \vec{D}^{T}_{s} &=& \vec{t}_{s}^{\ T} \cdot \mathbf{M}_{s}.
    \label{eqn:DDef}
\end{eqnarray}
With these four sets of collective variables we can 
turn to the calculation of forces, which as noted 
in the main text, requires the calculation of gradients
with respect to the cartesian coordinates of atom positions.
    
\subsection{Calculus of rotations}
    
The calculated rigid base-pair coordinates fall into two  
different categories: translations and rotations. 
For the two sets of translations derivation and thus 
the application of forces is simple, the two sets of 
rotations require more work. 
Here the construction of gradients of those two sets 
of rotations $\Vec{\Omega}$ and $\Vec{\omega}$ to small 
rotations of the underlying base frames 
$\text{d}{\vec{\phi}_{b}}$ is discussed.
    
To start, the derivative of the base reference frame 
$\mathbf{B}_{b}$ to rotations $\text{d}{\vec{\phi}^{b}}$ 
about its own axis can be found as follows.
Subsequently imposing the infinitesimal rotations 
$\text{d}{\phi}_{b}^{0}, \text{d}{\phi}_{b}^{1}, 
\text{d}{\phi}_{b}^{2}$ on $\mathbf{B}_{b}$ yields
\begin{equation*}
    \mathbf{B}_{b}+\text{d}{\mathbf{B}_{b}} = 
    \mathbf{B}_{b}\circ  
    \begin{pmatrix}
        1 & -\text{d}\phi^{3} & \text{d}\phi^{2} \\
        \text{d}\phi^{3} & 1 & -\text{d}\phi^{1} \\
        -\text{d}\phi^{2} & \text{d}\phi^{1} & 1 
\end{pmatrix}\\
\label{eqn:BaseDer}
\end{equation*}
   Which implies the derivative of $\mathbf{B}^{b}$ is given by
    \begin{equation}
        \dfrac{\text{d}{\mathbf{B}_{b}}}
        {\text{d}{\vec{\phi}_{b}}} = \mathbf{B}_{b}\Vec{\mathbf{S}}.
    \end{equation}
    Where $\Vec{\mathbf{S}}$ is a vector of skew matrices defined by
    \begin{equation*}
        \Vec{\mathbf{S}} = \left[ \begin{pmatrix}
0& 0 & 0 \\
0 & 0 & -1\\
0 & 1 & 0 
\end{pmatrix}, 
\begin{pmatrix}
0 & 0 & 1 \\
0 & 0 & 0 \\
-1 & 0 & 0
\end{pmatrix},
\begin{pmatrix}
0 & -1 & 0 \\
1 & 0 & 0 \\
0 & 0 & 0
\end{pmatrix} \right]^{T}.
    \end{equation*}
    
From this result the derivatives of the other relevant rotation 
matrices used above can be calculated. To start the derivatives 
of the $\mathbf{R}^{u}$ are given by:
\begin{equation}
    \dfrac{\partial{\mathbf{R}_{u}}}{\partial{\vec{\phi}_{ul}}} = 
    \mathbf{B}_{ur}\vec{\mathbf{S}}^{T}\mathbf{B}_{ul}^{T}, \quad 
    \dfrac{\partial{\mathbf{R}_{u}}}{\partial{\vec{\phi}_{ur}}} = 
    \mathbf{B}_{ur}\vec{\mathbf{S}}\mathbf{B}_{ul}^{T}.
\end{equation}
For the mid-pair frame $\mathbf{M}_u$ we find:
\begin{equation}
    \dfrac{\partial{\mathbf{M}_{u}}}{\partial{\vec{\phi}_{ur}}} = 
    \dfrac{\partial{\sqrt{\mathbf{R}_{u}}}} {\partial{\vec{\phi}_{ur}}} 
    \mathbf{B}_{ul} = \frac{1}{2}\sqrt{\mathbf{R}_{u}} \mathbf{R}_{u}^{T} 
    \dfrac{\partial{\mathbf{R}_{u}}}{\partial{\vec{\phi}_{ur}}} 
    \mathbf{B}_{ul}\\
    = \frac{1}{2} \mathbf{M}_{u}\vec{\mathbf{S}},
\end{equation}
and likewise:
\begin{equation}
    \dfrac{\partial{\mathbf{M}_{u}}}{\partial{\vec{\phi}_{ul}}}
    = \frac{1}{2} \mathbf{M}_{u}\vec{\mathbf{S}}.
\end{equation}
Where we used the fact that $\mathbf{R}_{u}$, $\sqrt{\mathbf{R}_{u}}$, 
$\mathbf{R}_{u}^{T}$, and  $\sqrt{\mathbf{R}_{u}^{T}}$ all commute and 
we used both representations of the midframe in Eq.~\eqref{eqn:midpoint}.
Similar results can be obtained for the lower basepair in a given 
base-pair step. 
As such the derivatives of the rotation matrices $\mathbf{R}_{s}$ and 
$\mathbf{M}_{s}$ relevant for the step parameters can also be written down.
Doing this gives for the rotation:
\begin{eqnarray}
    \dfrac{\partial{\mathbf{R}_{s}}}{\partial{\vec{\phi}_{ul}}} = 
    \dfrac{\partial{\mathbf{R}_{s}}}{\partial{\vec{\phi}_{ur}}} = \frac{1}{2} 
    \mathbf{M}_{u}\vec{\mathbf{S}}\mathbf{M}_{d}^{T}, \\
    \dfrac{\partial{\mathbf{R}_{s}}}{\partial{\vec{\phi}_{dl}}} = 
    \dfrac{\partial{\mathbf{R}_{s}}}{\partial{\vec{\phi}_{dr}}} = 
    \frac{1}{2}\mathbf{M}_{u}\vec{\mathbf{S}}^{T}\mathbf{M}_{d}^{T}.
\end{eqnarray}
where it should be noted derivatives to both base in the same pair have 
become equal. Likewise for the midframe of the step we find:
\begin{equation}
    \dfrac{\partial{\mathbf{M}_{s}}}{\partial{\vec{\phi}_{ul}}}
    = \dfrac{\partial{\mathbf{M}_{s}}}{\partial{\vec{\phi}_{ur}}}
    = \dfrac{\partial{\mathbf{M}_{s}}}{\partial{\vec{\phi}_{dl}}}
    = \dfrac{\partial{\mathbf{M}_{s}}}{\partial{\vec{\phi}_{dr}}}
    = \frac{1}{4} \mathbf{M}_{s}\vec{\mathbf{S}}.
\end{equation}
Now we have calculated the derivatives of all rotation matrices and 
$\mathbf{R}$ and mid frames $\mathbf{M}$ to small rotations of the bases. 
Subsequently we need to compute the derivatives of axis angle coordinates 
to the corresponding rotation matrix $\mathbf{R}$. To do this one can 
simply refer to the formulas to construct the axis angle representation 
from a matrix in \eqref{eqn:MatrixToAngle} and \eqref{eqn:MatrixToAxis}, and 
derive them:
\begin{eqnarray}
    \dfrac{\partial{\Theta(\mathbf{R})}}{\partial{\mathbf{R}}} 
    &=& \frac{-1}{2\sin{\Theta(\mathbf{R})}} \mathbf{I}_{3x3},\\
    \dfrac{\partial{\hat{K}(\mathbf{R})}}{\partial{\mathbf{R}}} 
    &=& \frac{\vec{\mathbf{S}} - \cot{\Theta(\mathbf{R})}\hat{K}(\mathbf{R})
    \otimes\mathbf{I}_{3x3}} {2\sin{\Theta(\mathbf{R})}}.
\end{eqnarray}
With this final bit of information the derivative of any one of our 
angular coordinates to a small rotation of a base frame can be written. 
To do this one needs only the relevant formula \eqref{eqn:OmegaDef} and 
the equations derived in this section. For example deriving some step 
rotation $\alpha$ to a rotation of base $b$ about axis $k$ would read:
\begin{align}
    \dfrac{\partial \Omega_{s}^{\alpha} }{\partial \phi^{k}_{b}} &= 
    \dfrac{\partial{\Theta_{s}}}{\partial{\phi^{k}_{b}}}\left[\hat{K}_{s}
    \cdot \mathbf{M}_{s}[\alpha]\right] \nonumber\\
    &+ \Theta_{s}\left[ \dfrac{\partial{\hat{K}_{s}}}{\partial{\phi^{k}_{b}}}
    \cdot \mathbf{M}_{s}[\alpha]  +  \hat{K}_{s}\cdot 
    \dfrac{\partial{\mathbf{M}_{s}[\alpha]}}{\partial{\phi^{k}_{b}}}\right].
\end{align}
Here we dropped the explicit indication that $\Theta_{s}$ and $\hat{K}_s$ 
are functions of $\mathbf{R}_s$. Their derivatives are calculated using 
the chain rule:
\begin{equation}
    \dfrac{\partial{\Theta_{s}}}{\partial{\phi^{k}_{b}}} = 
    \dfrac{\partial{\Theta_{s}}}{\partial{\mathbf{R}_{s}}}
    \dfrac{\partial{\mathbf{R}_{s}}}{\partial{ \phi^{k}_{b}}}, 
    \quad 
    \dfrac{\partial{\hat{K}_{s}}}{\partial{\phi^{k}_{b}}} = 
    \dfrac{\partial{\hat{K}_{s}}}{\partial{\mathbf{R}_{s}}}
    \dfrac{\partial{\mathbf{R}_{s}}}{\partial{ \phi^{k}_{b}}}.
\end{equation}
Subsequently these derivatives can be combined into the gradient 
of a rotational rigid base coordinate to the orientation of an 
underlying base by:
\begin{equation}
    \vec{\nabla}_{b}^{r} \Omega_s^{\alpha} = 
    \sum_{k} \dfrac{\partial \Omega_{s}^{\alpha} }{\partial \phi_b^{k}} 
    \mathbf{B}_{b}[k].
\end{equation}
Note that replacing $\Omega$ by $\omega$ and $s$ by $p$ in the previous 
three equation would yield the equivalent formulas for a pair rotation. 
        
\subsection{Rigid body dynamics}
    
Having calculated the gradients of the desired collective variables, 
we can recall Eq.~\eqref{Fc_simple}. While the precise gradients 
to atomic positions $\vec{A}_{i}$ are not known, and thus we can not 
write out equations for the atomic forces, the rotational gradients
to small rotations of the bases are known. Considering the bases as rigid 
we can then write out an equation for the torque $\vec{\tau}_{b}$ 
applied by some potential $V(X^{\alpha})$ on a base $\mathcal{B}_{b}$
\begin{equation}
    \Vec{\tau}_{b} = - \dfrac{\partial V(X^{\alpha})}{\partial X^{\alpha}} 
    \vec{\nabla}_{b}^{r}X^{\alpha}.
    \label{eqn:TorquePotDefinition}
\end{equation}
Here the torque is given in coordinates with respect to the lab frame
(the reference frame in which the atomic coordinates are provided by 
the MD engine). In order to have this torque act during an MD simulation 
it needs to be converted to a force for every atom $i$. 
The needed force can be found as follows
    \begin{equation}
        \vec{F}_{i} = m_{i}\vec{a}_{i} = m_{i}[ \vec{r}_{i} \times \vec{\alpha}_{b}] = 
        m_{i} \frac{\vec{r}_{i} \times \vec{\tau}_{b}} {I_{b}}.
        \label{eqn:forceFromTorque}
    \end{equation}
Here $\vec{r}_i$ is the vector pointing from the center of mass 
of the base $\vec{r}_{b}^{com}$ to the position $\vec{A}_i$, of atom $i$
$\vec{r}_{i}^{T} = [\vec{A}_i - \vec{r}_{b}^{com}]^{T}$. Additionally 
$I_{b}$ has been used to represent the moment of inertia of base with 
respect to the axis of the torque. Note that as the axis of the torque 
changes every time the constraint forces need to be calculated, so 
does $I_{b}$. Additionally $I_{b}$ will change due to fluctuations 
within the base.
    
Aside from the rotations $\vec{\omega}$ and $\vec{\Omega}$ one might 
also want to impose some bias on the translations $\vec{d}$ and $\vec{D}$. 
These are essentially the displacement between two rigid bodies expressed 
in the relevant midframe $\mathbf{M}$, and as such the gradients
of these to atom position $\vec{A}_i$ are 
\begin{eqnarray}
    \vec{\nabla}_{i}d^{\alpha}&=& \pm\frac{m_{i}}{m_{base}}
    \mathbf{M}_{pair}[\alpha], \label{eqn:Sder}\\
    \vec{\nabla}_{i}D^{\alpha}&=& \pm\frac{m_{i}}{m_{pair}}
    \mathbf{M}_{step}[\alpha].\label{eqn:Dder}
\end{eqnarray}
Where the plus is used for the base above the step (noted ul and ur in 
the diagram of Fig.~\ref{fig:stepDiagram}) and for the bases on the righthand 
strand (noted dr and ur in the diagram of Fig.~\ref{fig:stepDiagram}) 
when considering $\vec{D}$ and $\vec{S}$ respectively.
    
Finally equations \eqref{eqn:TorquePotDefinition} and \eqref{eqn:forceFromTorque} 
can be combined with equations \eqref{eqn:Sder}, \eqref{eqn:Dder} and 
\eqref{Fc_simple} to give the final force on an atom $i$ in base 
$b$ (and thus in pair $p=b/2$ and steps $s=b/2$ and $s=b/2 - 1$) due to a 
potential on the collective variables $V(\vec{\omega}_{p}, \vec{\Omega}_{s}, 
\vec{d}_{p}, \vec{D}_{s})$
\begin{align}
    \vec{F}_{i} = &-\sum_{\alpha} \dfrac{\partial V}
    {\partial \Omega^{\alpha}_{\frac{b}{2}}} 
    \frac{m_{i}}{I_{b}}\left[\vec{r}_{i}\times \vec{\nabla}_{b}^{r}
    \Omega^{\alpha}_{\frac{b}{2}}\right]\nonumber\\
    &-\sum_{\alpha}\dfrac{\partial V}
    {\partial D^{\alpha}_{\frac{b}{2}}}\vec{\nabla}_{i}
    D^{\alpha}_{\frac{b}{2}}\nonumber\\
    &-\sum_{\alpha} \dfrac{\partial V}{\partial \omega^{\alpha}_{\frac{b}{2}}} 
    \frac{m_{i}}{I_{b}}\left[\vec{r}_{i}\times 
    \vec{\nabla}_{b}^{r}\omega^{\alpha}_{\frac{b}{2}}\right]\nonumber\\
    &-\sum_{\alpha}\dfrac{\partial V}{\partial d^{\alpha}_{\frac{b}{2}}}
    \vec{\nabla}_{i}d^{\alpha}_{\frac{b}{2}}\nonumber\\
    &-\sum_{\alpha} \dfrac{\partial V}{\partial \Omega^{\alpha}_{\frac{b}{2}-1}} 
    \frac{m_{i}}{I_{b}}\left[\vec{r}_{i}\times \vec{\nabla}_{b}^{r}
    \Omega^{\alpha}_{\frac{b}{2}-1}\right]\nonumber\\
    &-\sum_{\alpha}\dfrac{\partial V}{\partial D^{\alpha}_{\frac{b}{2}-1}}
    \vec{\nabla}_{i}D^{\alpha}_{\frac{b}{2}-1}.
\end{align}
For completeness we note that with respect to \eqref{Fc_simple} we have replaced 
the exact positional gradients with versions that apply rigid body dynamics.
For translations they are replaced with gradients weighed to ensure all atoms in a base 
move together. For rotations we have substituted $\frac{m_i}{I_b}\left[\vec{r}_{i} 
\times \vec{\nabla}^{r}_{b}\right]$ for $\vec{\nabla}_{i}$. This last substitution 
is equivalent to projecting $\vec{\nabla}_{i}$ onto the plane normal to $\vec{r}$ 
and again reweighing which ensures that constraint forces do not distort 
the internal structure of the bases.

\section{Biasing Twist and Roll}

To explain the response of the DNA under the effect of a bias in the 
twist and in the roll, we consider the following free energy model
\begin{eqnarray}
f(\Omega_2,\Omega_3) &=& \frac 1 2 \left\{
A_2 \Omega_2^2 + C \Omega_3^2 + 2G \Omega_2 \Omega_3 + 
\right.
\nonumber\\
&& \left.
K_r (\Omega_2-\bar{\Omega}_2)^2 +
K_t (\Omega_3-\bar{\Omega}_3)^2 
\right\}
\label{SI:free_en}
\end{eqnarray}
with $\Omega_2$ and $\Omega_3$ excess roll and twist and $A_2$, $C$ and $G$ 
the roll stiffness, the twist stiffness and the twist-roll coupling, respectively. 
Equation~\eqref{SI:free_en} ignores higher order anharmonic terms which
become important in the high deformation regime. The stiffnesses $A_2$, $C$ and $G$
are sequence-dependent \cite{lank03}. Here we consider the base pair step considered 
in Fig.~\ref{fig:DeformAndCouple}, which is the central AT step of a DD sequence.
The average excess roll and twist, $\langle \Omega_2 \rangle$ and 
$\langle \Omega_3 \rangle$ are obtained from the minimization of 
Eq.~\eqref{SI:free_en}.

We consider the two cases of a twist and roll bias.
A twist bias corresponds to $K_r=0$ and $K_t \neq 0$ in \eqref{SI:free_en}.
The minimum is obtained by setting the partial derivatives of the free energy $f$ 
with respect to $\Omega_2$ and $\Omega_3$ to zero. In particular the condition
$\partial f/\partial \Omega_2=0$ gives
\begin{equation}
    \langle \Omega_3 \rangle = -\frac{A_2}{G} \langle \Omega_2 \rangle
    \label{SI:twist}
\end{equation}
In the case of a roll bias, corresponding to $K_r \neq 0$ and $K_t = 0$
in \eqref{SI:free_en}, the condition $\partial f/\partial \Omega_3=0$ gives
\begin{equation}
    \langle \Omega_3 \rangle = -\frac{G}{C} \langle \Omega_2 \rangle
    \label{SI:roll}
\end{equation}

Equations \eqref{SI:twist} and \eqref{SI:roll} give a linear dependence of 
$\langle \Omega_3 \rangle$ vs. $\langle \Omega_2 \rangle$. From prior work 
we know that $G >0$ \cite{skor18}, which implies that $\langle \Omega_3 \rangle$ vs. 
$\langle \Omega_2 \rangle$ lines must have negative slopes. This is agreement 
with the simulation data of Fig.~3(b). Note that the two slopes are different,
which is in agreement with \eqref{SI:twist} and \eqref{SI:roll}. On average 
$G$ is smaller than the two other couplings $A_2$ and $C$, which suggests a
large slope for a twist bias \eqref{SI:twist} and a small slope for a roll
bias \eqref{SI:roll}, which is indeed observed in the simulation data of 
Fig.~3(b).

\end{document}